# A 65nm 8b-Activation 8b-Weight SRAM-Based Charge-Domain Computing-in-Memory Macro Using A Fully-Parallel Analog Adder Network and A Single-ADC Interface

Guodong Yin[1], Mufeng Zhou[1], Yiming Chen[1], Wenjun Tang[1], Zekun Yang[1], Mingyen Lee[1], Xirui Du[1], Jinshan Yue[2], Jiaxin Liu[3], Huazhong Yang[1], Yongpan Liu[1], Xueqing Li[1]

[1]BNRist/ICFC, EE Dept., Tsinghua University, Beijing, China; [2]Institute of Microelectronics of the Chinese Academy of Sciences, Beijing, China; [3]University of Electronic Science and Technology of China, Chengdu, China

Contact: xueqingli@tsinghua.edu.cn

***Abstract*—Performing data-intensive tasks in the von Neumann architecture is challenging to achieve both high performance and power efficiency due to the memory wall bottleneck. Computing-in-memory (CiM) is a promising mitigation approach by enabling parallel in-situ multiply-accumulate (MAC) operations within the memory with support from the peripheral interface and datapath. SRAM-based charge-domain CiM (CD-CiM) has shown its potential of enhanced power efficiency and computing accuracy. However, existing SRAM-based CD-CiM faces scaling challenges to meet the throughput requirement of high-performance multi-bit-quantization applications. This paper presents an SRAM-based high-throughput ReLU-optimized CD-CiM macro. It is capable of completing MAC and ReLU of two signed 8b vectors in one CiM cycle with only one A/D conversion. Along with non-linearity compensation for the analog computing and A/D conversion interfaces, this work achieves 51.2GOPS throughput and 10.3TOPS/W energy efficiency, while showing 88.6% accuracy in the CIFAR-10 dataset.**

## I. Introduction

To overcome the memory wall bottleneck, computing-in-memory (CiM) has been proposed to enable parallel and in-situ data processing within the memory with support from the peripheral interface and datapath [1]-[6]. Recently, SRAM-based charge-domain CiM (CD-CiM) has shown its potential of enhanced power efficiency and computing accuracy [1]-[3].

However, as shown in Fig. 1, existing SRAM-based CD-CiM faces severe scaling challenges to meet the throughput requirement of high-performance multi-bit-quantization applications. One cause is the scaling unfriendliness of exponentially increasing size of binary-weighted capacitors connected to the SRAM output bitlines for summing-up the computing results of multiple weight bits [1]. This becomes more challenging with ≥8bit quantization bits, as shown in Fig. 1(a). Another more critical cause is the A/D-related bottleneck: in conventional SRAM-based CD-CiM, the bit-serial activation input method needs an A/D converter (ADC) for the CiM output under each activation bit [3]. This A/D conversion needs to be accurate so as to avoid excessive distortion in the subsequent shift-and-add of the CiM outputs, resulting in immense ADC-related overheads during the trade-off between power, area, and latency [2][8]. Consequently, the existing SRAM-based CD-CiM throughput performance is limited by the reluctant ADC sharing, activation input bits splitting, and the number of simultaneously turned-on rows.

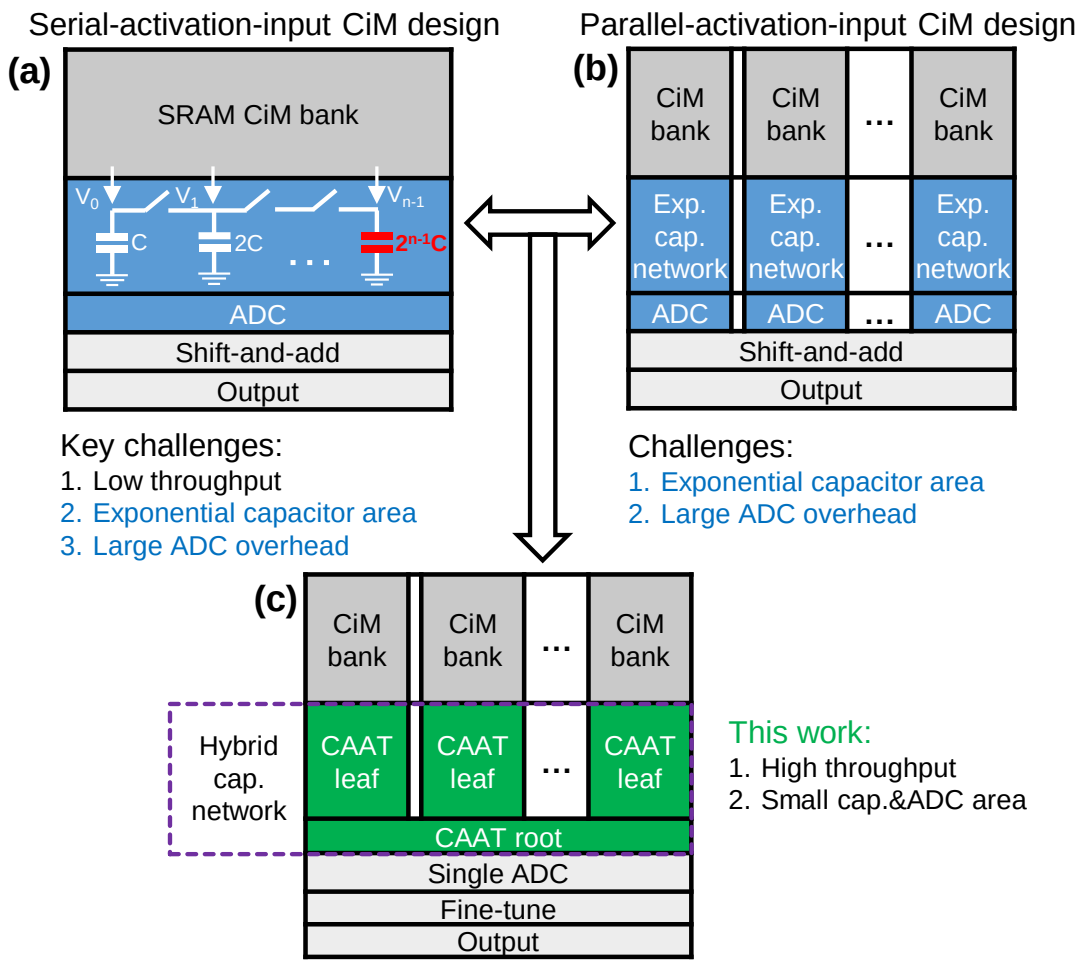

Fig. 1. Challenges in high-throughput multi-bit SRAM-based CD-CiM.

This work investigates these bottlenecks, and attempts to provide new approaches to SRAM-based high-throughput multi-bit-quantization CiM. For the aforementioned first cause, it is noted that area-efficient C-2C structure in existing ADC design could be borrowed for the CiM result sensing to reduce the capacitor-based summation area overheads. For the second cause related to ADC overheads, it is noted that there exists one intuitive approach of providing multiple replicated CiM banks for different activation input bits, namely the parallel-activation-input CiM design in Fig. 1(b). However, the area and power costs per activation bit remains unchanged. Can we reduce the overheads with parallel activation bits?

This work addresses this question and makes contribution shown in Fig. 1(c): a high-throughput SRAM-based ReLU-optimized CD-CiM architecture enabled by a two-level compact charge-domain analog adder tree (CAAT) and ReLU-optimized one-ADC sensing interface for the entire CiM array of all parallel CiM banks. It could complete MAC+ReLU computing of two signed 8b vectors in one CiM cycle with one-time A/D conversion, leading to 8x ADC-related area and energy savings compared with prior works. To ensure the computing accuracy, a non-linearity fine-tune scheme is also proposed to compensate the CAAT and ADC non-linearities by tuning the mean and variation values of the capacitor-based summation network coefficients.

In the rest of this paper, Section II presents the proposed CiM architecture and design details. Section III shows and discusses the measurement results of the macro chip fabricated in 65nm CMOS. Section IV concludes this work.

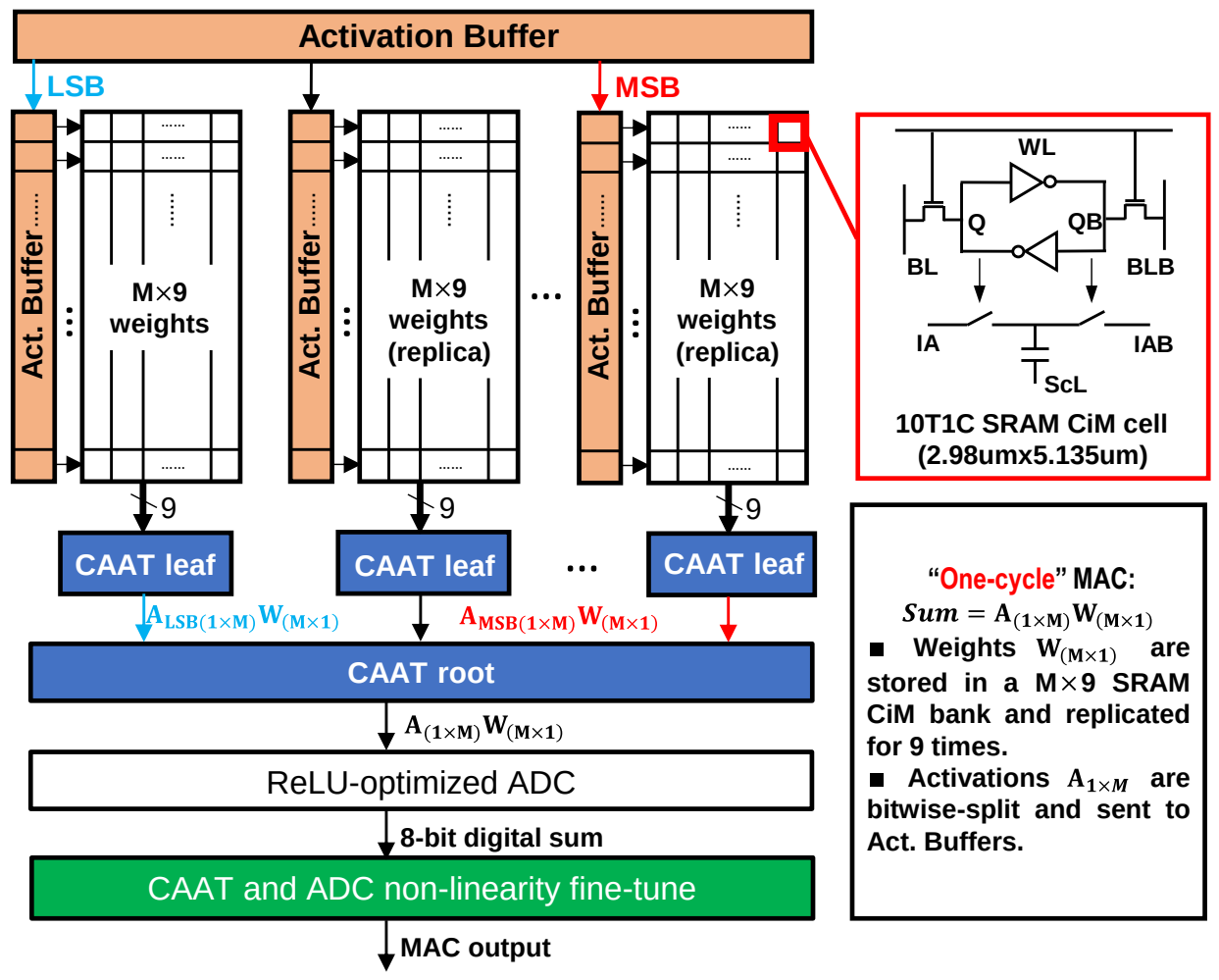

Fig. 2. Challenges for high-throughput multi-bit SRAM-based CD-CiM.

## II. Proposed Charge-Domain SRAM-Based CiM

### *A. Overall Architecture*

Fig. 2 depicts the overall architecture of the proposed SRAM-based CD-CiM macro. Different from the parallel-activation-input approach in Fig. 1(b) that uses replicated CiM banks with binary-weighted capacitor summing network and a dedicated ADC for each bank, this work proposes a two-level hybrid binary-C-2C compact charge-domain analog adder tree (CAAT) followed by one single ADC for the entire CiM array of all banks to achieve fast speed (one-CiM-cycle latency), light-weight summation network (rather than exponential binary-weighted capacitor network), and scaling-friendly results quantization (only one ADC instead of many).

The function of the proposed CiM macro could be expressed as Sum = $\mathbf{A}_{1\text{xM}}$ x $\mathbf{W}_{\text{Mx1}}$. Same as [2], this work uses N+1 bits to represent an N-bit number $x$:

$$x = \sum_{i}^{N-1} n_i \times 2^{i-1} + (n_{0+} + n_{0-}) \times 2^{-1}. \qquad (1)$$

Therefore, this work uses 9 bits to represent an 8b signed number. In addition to the peripherals, the proposed CiM macro includes: (i) an SRAM CiM array with 9 replicated 9-column CiM banks, each containing multiple rows of 10T1C SRAM CiM cells for computing with one bit of the signed 8-bit activation input; (ii) the proposed CAAT computes the weighted MAC summation results of a total of 81 outputs from 9 CiM banks within one CiM cycle; (iii) one ReLU-optimized ADC for the entire CiM array, featured with early-stop-to-zero when MSB is negative by definition of ReLU.

### *B. Charge-Domain Analog Adder Tree*

Fig. 3 illustrates the proposed area-efficient charge-domain analog adder tree (CAAT), and the 8b MAC operation in one CiM cycle. The CAAT includes CAAT root (CAAT-R) for the summation of outputs from 9 CAAT leaves (CAAT-L). Different from a fully binary-weighted capacitor network in prior works, this work uses a hybrid binary-C-2C capacitor network for analog summation. As a typical C-2C capacitance network has matching accuracy around 5-6 bits [7], this work does not use a C-2C structure for all 8 bits. Instead, the higher 4 bits in each CAAT-L are binary weighted, and the lower bits are implemented in a C-2C manner for reduced total capacitor size. In Fig. 3(a), to ensure computing accuracy, the load capacitor in each source line (ScL) is designed equal. To further reduce the capacitor network area, this work splits the MSB capacitor (16C) into two 8C capacitors. Therefore, the total load capacitor in each ScL is 9C (=8C+1C) rather than 17C (=16C+1C without splitting).

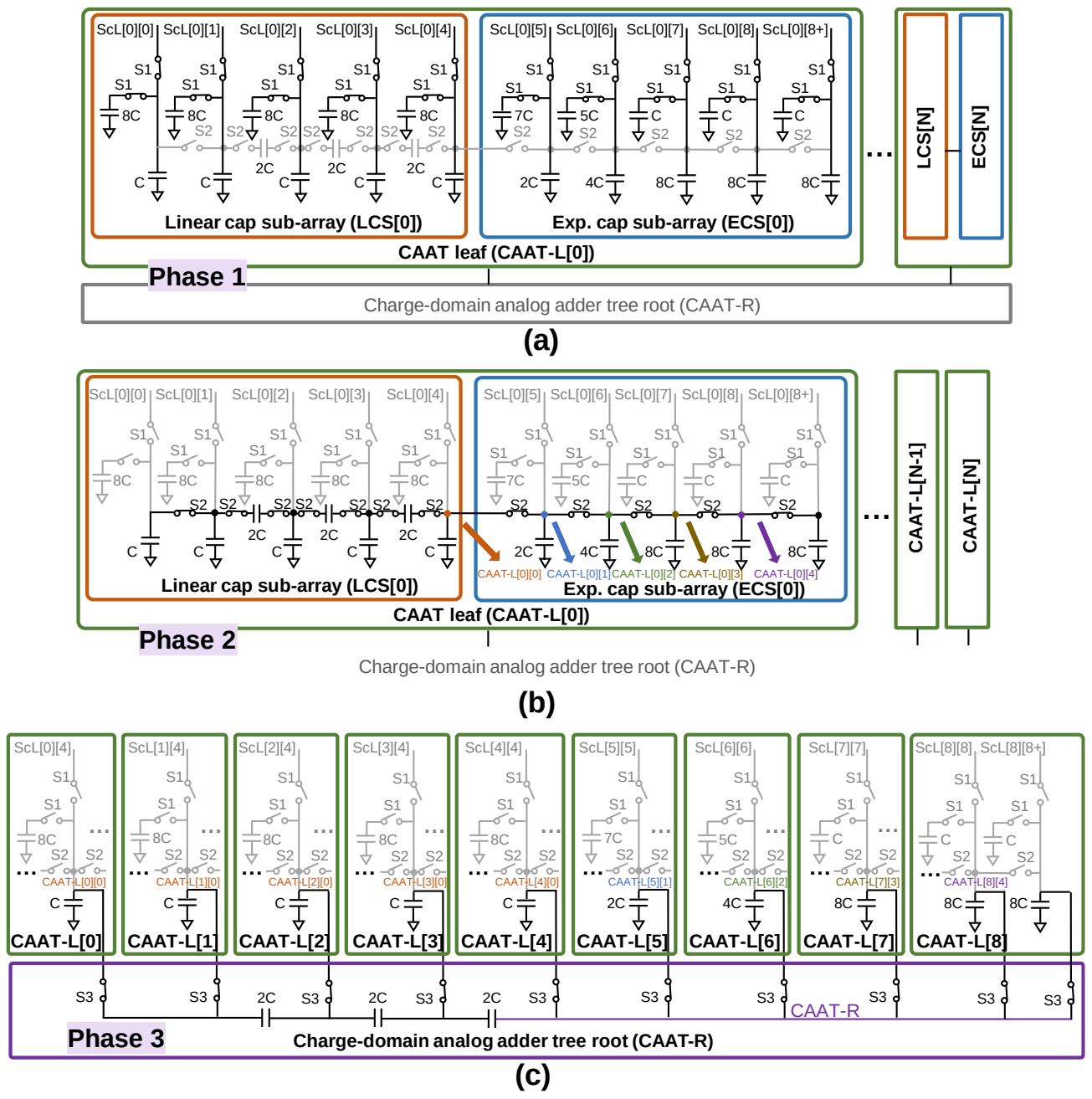

Fig. 3. Proposed charge-domain analog adder tree (CAAT).

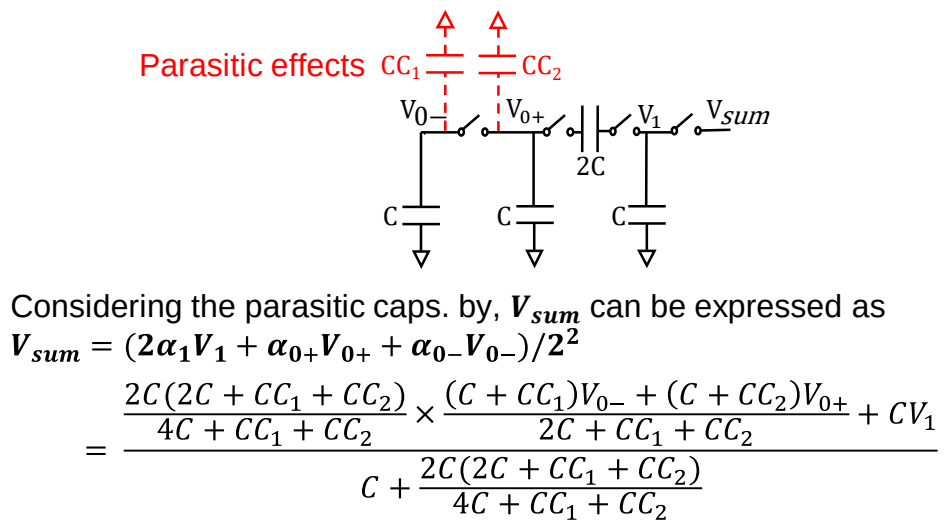

Fig. 4. An example of capacitor network non-linearity.

As shown in Fig. 3, the 8b MAC operation is performed in 3 phases: in-column summation, in-bank summation, and in-array summation. In the in-column summation phase, S1=ON and S2=S3=OFF, an averaged 1b MAC operation $\sum_j A_j[k] \times W_j[i]$ ($A_j[k]$ is $k^{th}$ bit of $j^{th}$ activation, and $W_j[i]$ is $i^{th}$ bit of $j^{th}$ weight) is carried out as CiM cells in the same column couple the charge onto the source line (ScL) with an equal load capacitor. In the in-bank summation phase, S2=ON and S1=S3=OFF. With the merged voltage nodes (CAAT-L[i][0:8]), the charge stored on different ScL capacitors will be redistributed to get the averaged summation result in the bank level. After the conduction of weight bits summation $\sum_j A_j[k] \times W_j$ in the first two phases, activation bits summation $\sum_j A_j \times W_j$ is performed in the third phase with S3=ON and S1=S2=OFF. CAAT-L outputs in different banks, i.e. CAAT-L[i], will be combined through CAAT-R, whose voltage is converted by the 8b ReLU-optimized ADC. The ADC overhead is greatly reduced because only one A/D conversion is needed for an 8b MAC operation.

Furthermore, as this design could finish the entire MAC operation in the analog domain through the CAAT, it is a natural thought to combine the ADC and the ReLU function in this design. When the MSB of the ADC output indicates a negative MAC result, there is no need to convert the other ADC output bits as the ReLU will reset all bits to zero. This optimization could reduce the ADC energy consumption by half. It is noted that such an optimization does not apply to the serial-activation-input works [1][2] where one A/D conversion is needed for each partial sum.

### C. Proposed Fine-Tune Compensation Scheme

While the computing results in this work rely on the accuracy of CAAT and ADC, the parasitic and other non-linear effects would introduce distortion in the computation result. Capacitor networks such as CAAT and the capacitive DAC in ADC suffer from parasitic effects and random mismatches, which are the main non-linearity sources in these circuits. In the context of CiM, the capacitor network output with deviations affects the inference accuracy. Fig. 4 presents an example in a capacitor network.

It is also noted that such non-linearity is mostly determined once the chip is fabricated, along with some drifting dynamics during use. While the calibration of many capacitors could be too costly, previous compensation efforts for the ADC non-linearity through the weight matrix tuning could also cause large retrain overheads to perform on each edge device, as shown in Fig. 5(a) [9].

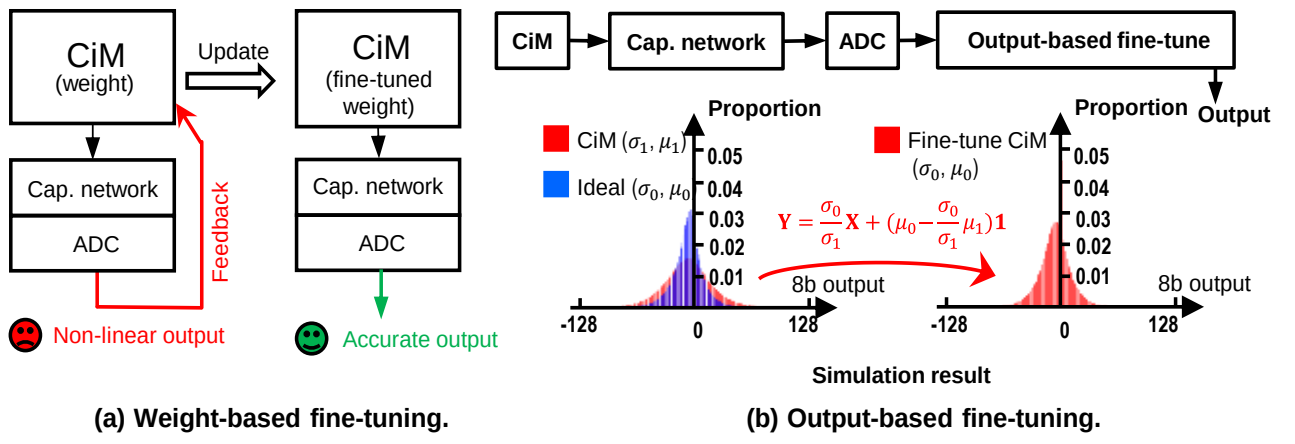

Fig. 5. Details of CAAT and ADC nonlinearity fine-tune.

Therefore, we provide a low-cost output-based fine-tune technique. Because the capacitor network is the main source of non-linearity, the distortion of the output can be modeled as a linear function, as shown in Fig. 5(b). Based on this observation, the aim of output-based fine-tune is to make the output activation have the same mean and variation as an ideal output activation. In detail, the ideal output is performed on software, while the inference on CiM is performed once to collect the ADC output. In this way, we can get the fine-tune parameters $\frac{\sigma_0}{\sigma_1}$ and $\mu_0 - (\frac{\sigma_0}{\sigma_1}\mu_1)$. The deployed neural network will adjust the mean and variance of the output based on these two parameters. The redundant inference on dataset is only performed once after the chip is taped out. After applying the output-based fine tuning, the simulated average has been improved significantly. The details are described in Section III.C. In addition, it is possible to tune the ADC output before it is used as the activation input to the next layer. Output-based fine-tune uses the ADC output vector statistics and adjusts its distribution dynamically. It could be a universal solution for different chips at the minor cost of extra hardware support for inference.

## III. Measurement And Evaluation

### A. Chip Fabrication and Measurement Setup

This work fabricated a test chip with a 10T1C macro using a 65nm CMOS technology. The SRAM-based CiM array includes 9 1152x9- banks. Fig. 6(a) presents the die photograph. Fig. 6(b) shows the test platform of the fabricated chip. The chip PCB, power source PCB, and test PCB are stacked on an FPGA board. The FPGA sends testbench of typical NN layers to the test chip and receives outputs of the test chip, which are then sent to a computer via the Ethernet.

### B. Energy, Latency, and Area Evaluation

Fig. 7 evaluates this work against the baseline design of parallel-activation-input CiM with exponential capacitor summing up networks. While both improve the overall throughput greatly, Fig. 7(a) reveals significant area cost differences. The capacitor area occupation of the baseline design increases rapidly as the bit width grows, while the proposed CAAT area increases much more slowly. For one 8b CAAT-L, the total capacitance is reduced by 10.8x, from 1032C to 96C.

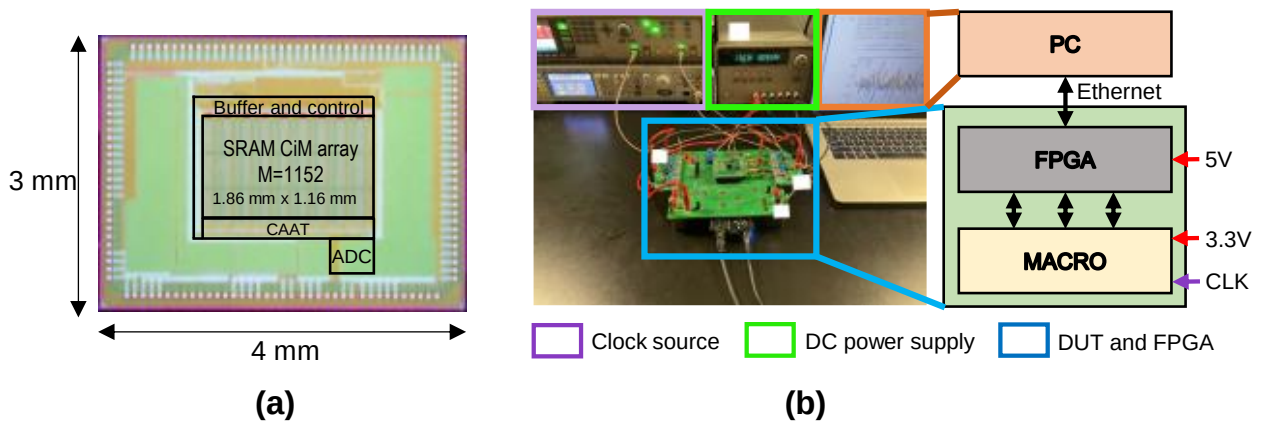

Fig. 6. Die photo and the test platform of the fabricated chip.

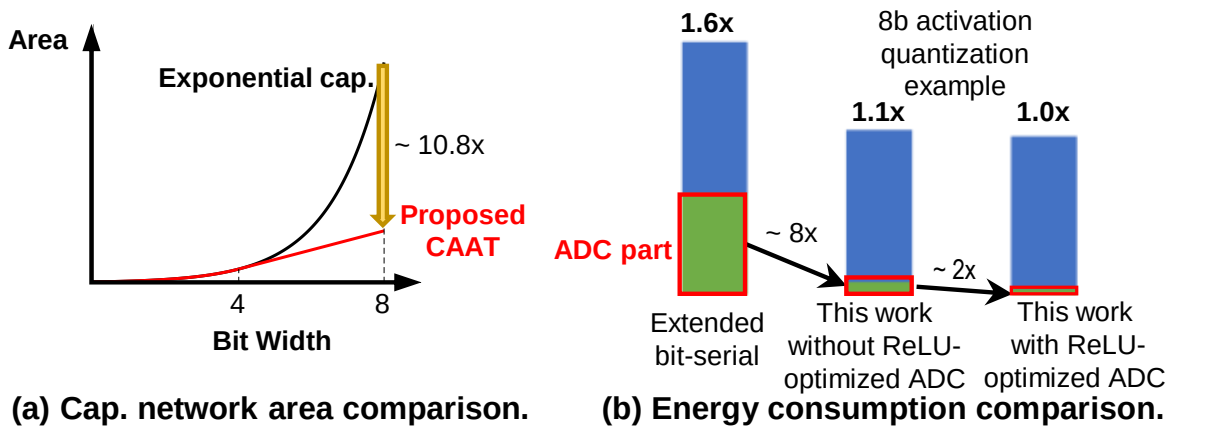

Fig. 7. Key design evaluations: (a) capacitor network area comparison; (b) energy consumption comparison.

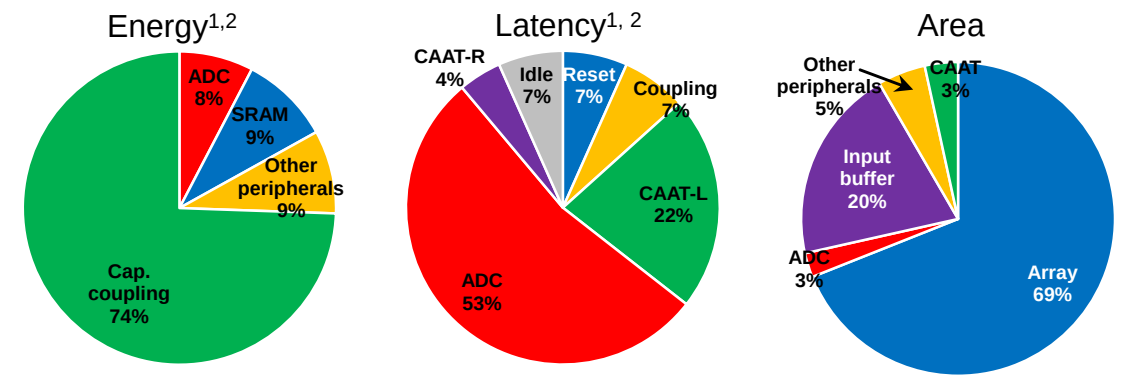

[1]: Simulated at 1.0V, 500MHz clock for ADC, 1.0V, 1GHz clock for others, random activations and weights.
[2]: Off-chip and IO power is not included. Off-chip operation time is not included.

Fig. 8. Energy, area, and latency breakdown of the proposed design.

Fig. 7(b) compares the energy consumption comparison. As the proposed CAAT design needs only one A/D conversion for one 8-bit MAC operation, the ADC energy consumption is about 1/8 of that in the parallel-activation-input CiM design. Further power reduction of ~2x is achievable using the ReLU-based early-stop-to-zero optimization when the MSB of the ADC output shows a negative output. At the macro level when peripherals are included, the proposed design has also achieved 1.6x improved energy efficiency, when compared with the parallel-activation-input CiM.

Compared with the parallel-activation-input CiM design with one-level capacitance summation (shown in Fig.1(b)), the latency is similar, and the total area occupation at the macro level has 1.2x savings, thanks to the hybrid binary-C-2C capacitor network.

TABLE I. COMPARISON TABLE WITH PREVIOUS CIM WORKS.

| | ISSCC'21 [1] | JSSC'20 [2] | ISSCC'21 [3] | ISSCC'20 [4] | ISSCC'21 [5] | ISSCC'20 [6] | **This Work** |
|---|---|---|---|---|---|---|---|
| **Technology** | 16nm | 65nm | 65nm | 28nm | 65nm | 28nm | **65nm** |
| **Power Supply(V)** | 0.8 | 1.2 | 1.1 | 0.7-0.9 | 1.0 | 0.85-1.0 | **0.76-1.2** |
| **Activation** | 1-8b | 1-8b | 8b | 4/8b | 2/4/6/8b | 8b | **8b** |
| **Weight** | 1-8b | 1-8b | 8b | 4/8b | 1-8b | 8b | **8b** |
| **ADC** | 8b | 8b | 8b | 5b | 0/2/4b | 5b | **8b** |
| **Clock Frequency (MHz)** | 200 | 100 | 50 | 200 | 25-100 | 200 | **240-1000** |
| **Throughput (GOPS)[1,4]** | N/A | 34.1 | 4.71 | 31.2 | N/A | N/A | **35.8[5]\|51.2[3]** |
| **Energy Efficiency (TOPS/W)[1]** | 30.2 | 3 | 4.76 | 16.6 | 3.66 | 7.3 | **10.1[5]\|3.53[3]** |
| **Energy Efficiency (TOPS/W)[1,2]** | 1.82 | 3 | 4.76 | 3.08 | 3.66 | 1.35 | **10.1[5]\|3.53[3]** |

[1]: 1 operation (OP) = 1 8b-8b addition/multiplication. Off-chip and IO power is not included; [2]: Scaled to 65nm, assuming energy $\propto$ (Tech)$^2$ [3]. [3]: Measured at 1.0V, 500MHz clock for ADC, 1.0V, 1GHz clock for others, 8b activation, 8b weight. [4]: Off-chip operation time is not included. [5]: Measured at 1.0V, 350MHz clock for ADC, 0.8V, 700MHz clock for others.

Fig. 8 shows the breakdown pie charts of energy, area, and latency. The energy and latency data are obtained through post-layout simulations under 1.0V supply, 500MHz clock for ADC, and 1.0GHz clock for others (with the same activations and weights as those used in the experimental measurements). The results do not include power for off-chip driver and IO, and do not include operation time by off-chip modules. It is shown that the ADC consumes only 3% area occupation and 8% energy consumption. Thanks to the single-ADC one-cycle-MAC architecture, the proposed design achieves the highest 51.2 GOPS throughput at 1.0 GHz clock frequency. Furthermore, at 240 MHz clock frequency, it achieves the highest 10.3 TOPS/W energy efficiency, thanks to less A/D conversions. These results indicate a promising architecture for high-throughput energy-efficient multi-bit SRAM-based CD-CiM. Table I compares the key metrics with some previous CiM works.

### C. *Precision Measurement and Analysis*

As mentioned in Section II.C, the computing accuracy depend on the linearity of CAAT and ADC. Fig. 9 illustrates the non-linearity of these sensitive circuit modules. Fig. 9(a) shows the simulated CAAT non-idealities in an integral non-linearity (INL) histogram based on post-layout simulations, indicating that ~70% CAAT samples have 7b summation accuracy. Fig. 9(b) plots the INL measurement results of the 8-bit ADC, showing a maximum INL of only 1.2 LSB, well meeting the requirement on the ADC quantization accuracy.

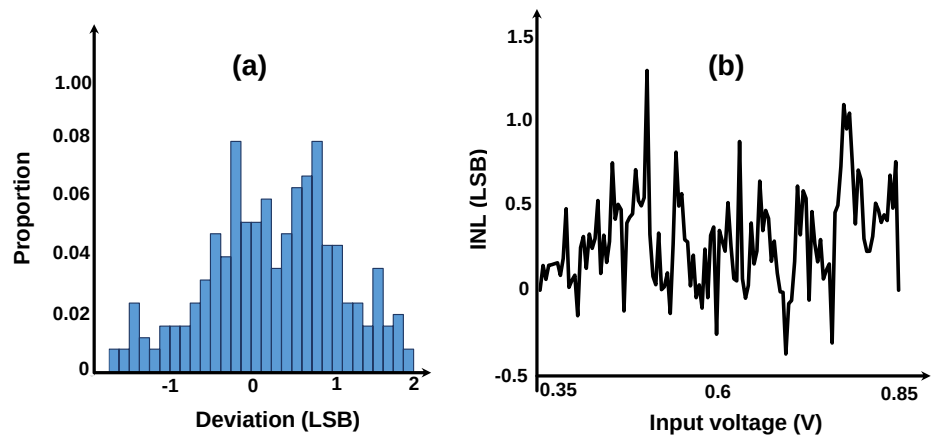

Fig. 9. Non-linearity of sensitive circuit modules: (a) CAAT post-layout analysis; (b) ADC measurement results.

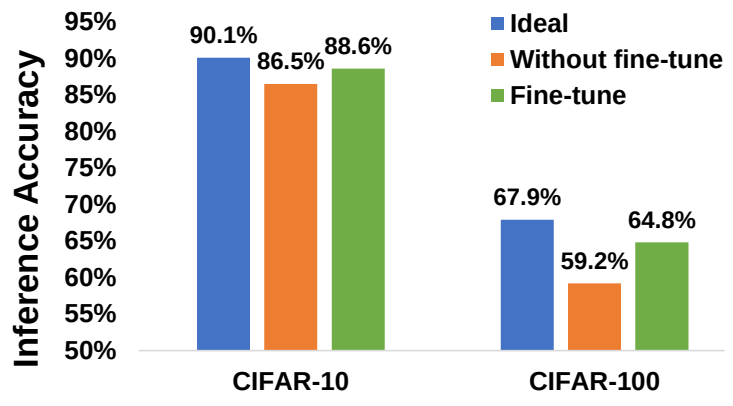

Fig. 10. Inference accuracy of the proposed SRAM-based CD-CiM.

In the context of inference accuracy, the adopted fine-tune method to reduce the effects of CAAT and ADC non-linearities has been evaluated based on the collected data in Fig. 9. Provided neural network models of 8b-input and 8b-weight VGG-8, simulation results show recovery of inference accuracy from 86.5% and 59.2% to 88.6% and 64.8% for CIFAR-10 and CIFAR-100 datasets, respectively, as shown in Fig. 10.

## IV. CONCLUSIONS

This paper has presented an SRAM-based single-ADC CD-CiM macro that could complete MAC+ReLU computing of two signed 8b vectors in one CiM cycle with only one-time A/D conversion. The analog weighted summing capacitor network (CAAT) is presented and evaluated, showing both high throughput and low area costs. A fine-tune method is proposed to reduce the non-linearity effect of parasitics and achieve high inference accuracy. The test chip has been fabricated in 65nm CMOS and measurements show that it achieves 51.2GOPS throughput, 10.3TOPS/W energy efficiency, and 88.6% accuracy in CIFAR-10 dataset.

## ACKNOWLEDGMENT

This work is supported in part by National Key R&D Program of China (#2019YFA0706100), in part by NSFC (#U21B2030, #61934005, #61720106013, #62174023), and in part by Beijing Nova Program (#Z211100002121125).

## REFERENCES

[1] H. Jia et al., "15.1 A Programmable Neural-Network Inference Accelerator Based on Scalable In-Memory Computing," in ISSCC 2021, pp. 236–238.

[2] H. Jia, H. Valavi, Y. Tang, J. Zhang, and N. Verma, "A Programmable Heterogeneous Microprocessor Based on Bit-Scalable In-Memory Computing," IEEE JSSC, vol. 55, no. 9, pp. 2609–2621, Sep. 2020.

[3] S. Xie et al., "16.2 eDRAM-CIM: Compute-In-Memory Design with Reconfigurable Embedded-Dynamic-Memory Array Realizing Adaptive Data Converters and Charge-Domain Computing," in ISSCC 2021, vol. 64, pp. 248–250.

[4] X. Si et al., "15.5 A 28nm 64Kb 6T SRAM Computing-in-Memory Macro with 8b MAC Operation for AI Edge Chips," in ISSCC 2020, pp. 246–248.

[5] J. Yue et al., "15.2 A 2.75-to-75.9TOPS/W Computing-in-Memory NN Processor Supporting Set-Associate Block-Wise Zero Skipping and Ping-Pong CIM with Simultaneous Computation and Weight Updating," in ISSCC 2021, vol. 64, pp. 238–240.

[6] J.-W. Su et al., "15.2 A 28nm 64Kb Inference-Training Two-Way Transpose Multibit 6T SRAM Compute-in-Memory Macro for AI Edge Chips," in ISSCC 2020, pp. 240–242.

[7] E. Alpman et al., "A 1.1V 50mW 2.5GS/s 7b Time-Interleaved C-2C SAR ADC in 45nm LP digital CMOS," in ISSCC 2009, pp. 76-77,77a.

[8] Jian-Wei Su et al., "16.3 A 28nm 384kb 6T-SRAM Computation-in-Memory Macro with 8b Precision for AI Edge Chips," in ISSCC 2021, pp. 250-252.

[9] S. Huang et al., "Mitigating Adversarial Attack for Compute-in-Memory Accelerator Utilizing On-chip Finetune," in 2021 IEEE 10th Non-Volatile Memory Systems and Applications Symposium, pp. 1-6.